\documentclass{moriond}

\def\be{\begin{equation}}
\def\ee{\end{equation}}
\def\bea{\begin{eqnarray}}
\def\eea{\end{eqnarray}}

\newcommand{\Photo}{}
\usepackage{lineno}

\begin{document}
\vspace*{4cm}
\title{Review of flavour physics at ATLAS and CMS}

\author{Anne-Mazarine Lyon, on behalf of the ATLAS and CMS Collaborations\footnote{Copyright 2026 CERN for the benefit of the ATLAS and CMS Collaborations. Reproduction of this article or parts of it is allowed as specified in the CC-BY-4.0 license.}}

\address{CERN, Esplanade des Particules 1\\
1217 Meyrin, Switzerland}

\maketitle\abstracts{
A review of recent results in flavour physics at the ATLAS and CMS experiments is presented. These include measurements of quarkonia and charm production cross sections, the B$^0$ lifetime, and the mass splittings between excited and ground B~meson states. Results targeting the characterisation of all-charm tetraquarks are also discussed. Finally,  studies of rare heavy-flavour decays are reported. The various analyses exploit Run~2 and partial Run~3 LHC data and contribute to pushing the precision frontier in flavour physics.
}

\section{Introduction}

Flavour physics is a very active research field, as it allows for precise tests of standard model (SM) processes and provides sensitive probes for new physics.  Over the past few years, the ATLAS~\cite{ATLAS} and CMS~\cite{CMS} experiments have established themselves as competitive facilities to study heavy-flavour decays, in a way complementary to flavour-dedicated experiments. This is made possible by their (i) nearly full solid-angle coverage, (ii) excellent vertexing performance, and (iii) outstanding muon reconstruction and identification capabilities. Additionally, the development of triggers selecting low-energy processes enhanced the acceptance for heavy-flavour decays~\cite{ATLAStriggers,CMStriggers}.

In this paper, we present a review of recent results in flavour physics at the ATLAS and CMS experiments.  Measurements of properties of heavy-flavour particles are discussed in Section~\ref{section_properties}. The state-of-the-art in the characterisation of all-charm tetraquarks and studies of rare heavy-flavour decays are presented in Sections~\ref{section_tetraquarks} and~\ref{section_rare_decays}, respectively.  The main conclusions are summarised in Section~\ref{section_summary}. The results are derived using LHC Run~2 and partial Run~3 proton-proton collision data, collected with triggers requiring at least two muons in the event.

\section{Measurements of heavy-flavour properties}\label{section_properties} 

\begin{figure}[tb!]
\centering
\raisebox{-0.5\height}{\includegraphics[width=0.43\textwidth]{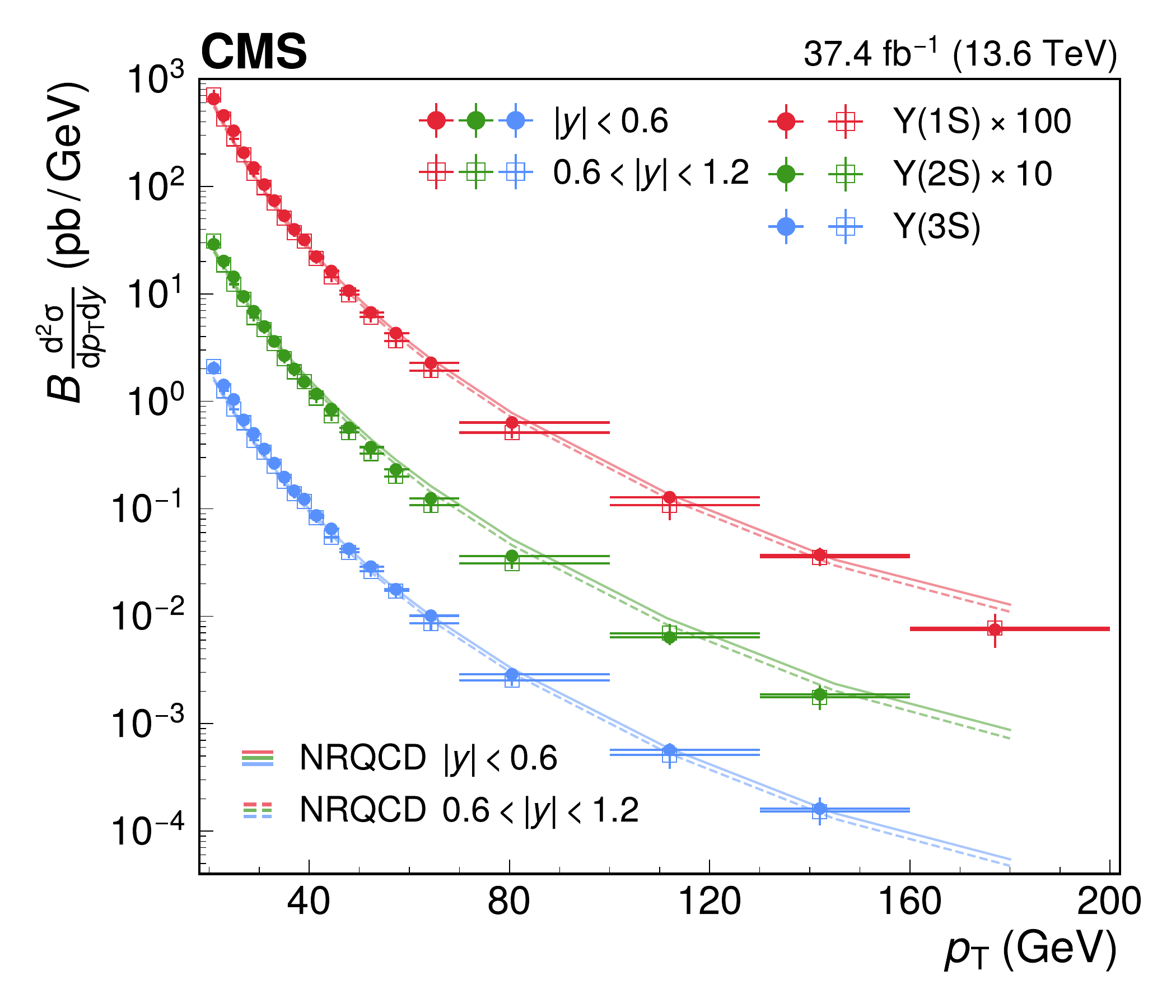}}
\hspace{1cm}
\raisebox{-0.5\height}{\includegraphics[width=0.39\textwidth]{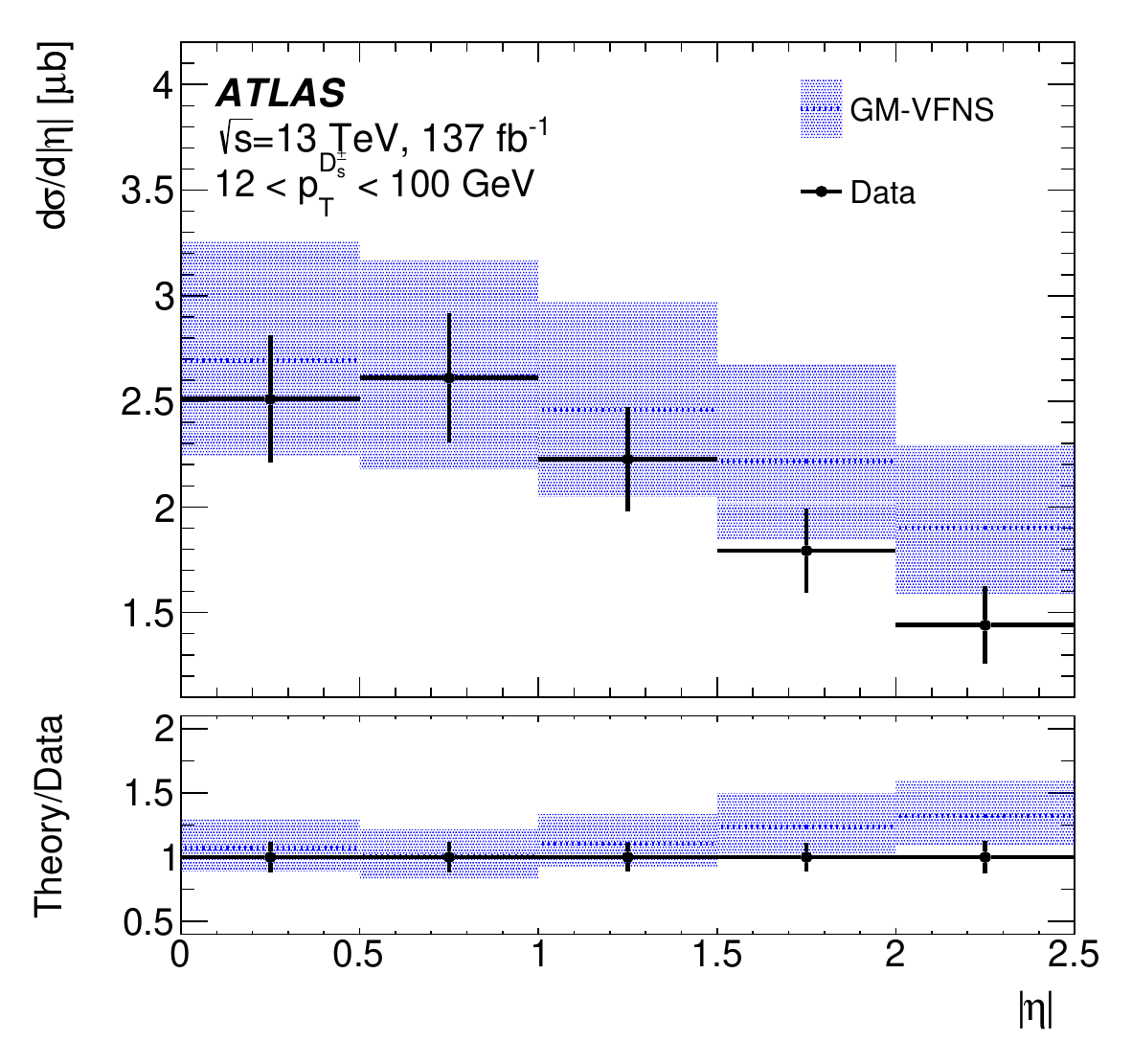}}
\caption{\textit{(Left)} Product of the differential cross section for the $\Upsilon$(1S), $\Upsilon$(2S), and $\Upsilon$(3S) states with the $\mu^+\mu^-$ decay branching fractions as a function of $p_\mathrm{T}$~\protect\cite{BPH-24-004}. The measurements are reported for two rapidity, $|y|$, ranges and compared with NRQCD predictions. The results for $\Upsilon$(1S) and $\Upsilon$(2S) are arbitrarily scaled to improve readability. \textit{(Right)} Differential cross sections of $\mathrm{D}_\mathrm{s}$ mesons as a function of $|\eta|$, for mesons with $12 < p_\mathrm{T} < 100$~GeV~\protect\cite{ATLAS2}. The results are compared with GM-VFNS predictions. \label{figure_a}}
\end{figure}

The cross sections for the production of $\Upsilon$($n$S), $n=1,\,2,\,3$, are measured at CMS for the first time at a centre-of-mass energy of 13.6~TeV~\cite{BPH-24-004}. Measuring the cross sections of these b$\overline{\mathrm{b}}$ states provides crucial insights into the hadron formation within the quantum chromodynamics (QCD)  framework. The analysis exploits Run~3 data (37.4~fb$^{-1}$), collected with the CMS B-parking dimuon triggers~\cite{CMStriggers}. The measurement is performed (i) differentially in the transverse momentum ($p_\mathrm{T}$) of the $\Upsilon$($n$S) meson, in a range extending for the first time to 200~GeV, and (ii) under the assumption that the meson is produced unpolarised.  The measured cross sections, multiplied by the branching fraction to dimuon decays, are shown in Fig.~\ref{figure_a} (left) and are found to be in good agreement with the non-relativistic QCD (NRQCD) theoretical predictions. 

A measurement of production cross sections of the D$^\pm$ and D$_\mathrm{s}^\pm$ mesons has recently been performed at ATLAS using Run~2 data (137~fb$^{-1}$)~\cite{ATLAS2}.  
The cross sections are measured differentially in bins of the D~meson  pseudorapidity, $|\eta|$, and $p_\mathrm{T}$, using the  $\mathrm{D}\to\phi(\mu^+\mu^-)\pi^\pm$ decay channel. For the first time, the measurement is extended up to $p_\mathrm{T} = 100$~GeV. Results are presented inclusively for prompt and non-prompt D~meson production. Figure~\ref{figure_a} (right) shows the cross sections for  D$_\mathrm{s}^\pm$ mesons in bins of $|\eta|$. The measurement agrees well with the general-mass variable-flavour-number scheme (GM-VFNS) theoretical predictions. 

\begin{figure}[b!]
\centering
\raisebox{-0.5\height}{\includegraphics[width=0.41\textwidth]{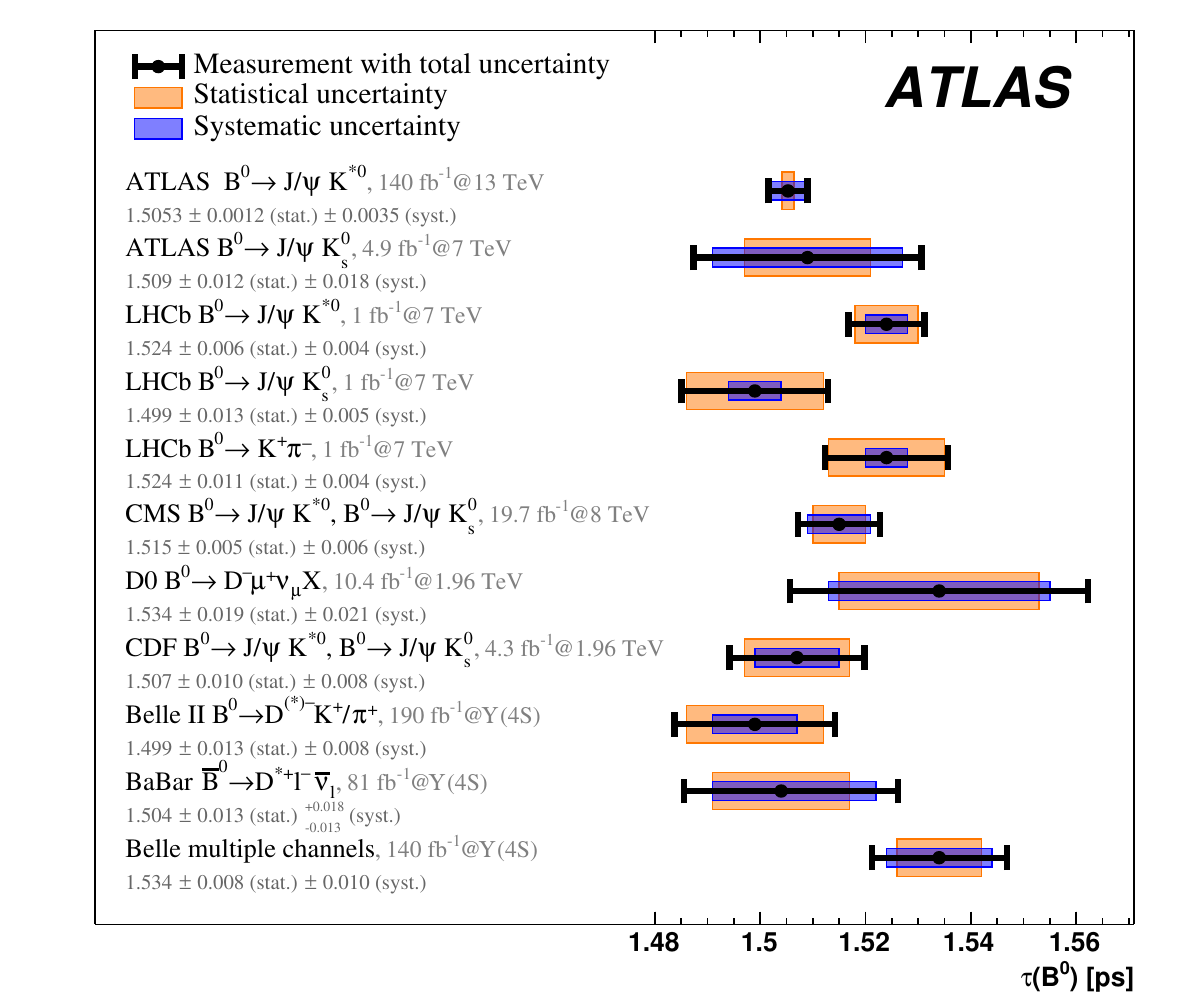}}
\hspace{0.7cm}
\raisebox{-0.5\height}{\includegraphics[width=0.42\textwidth]{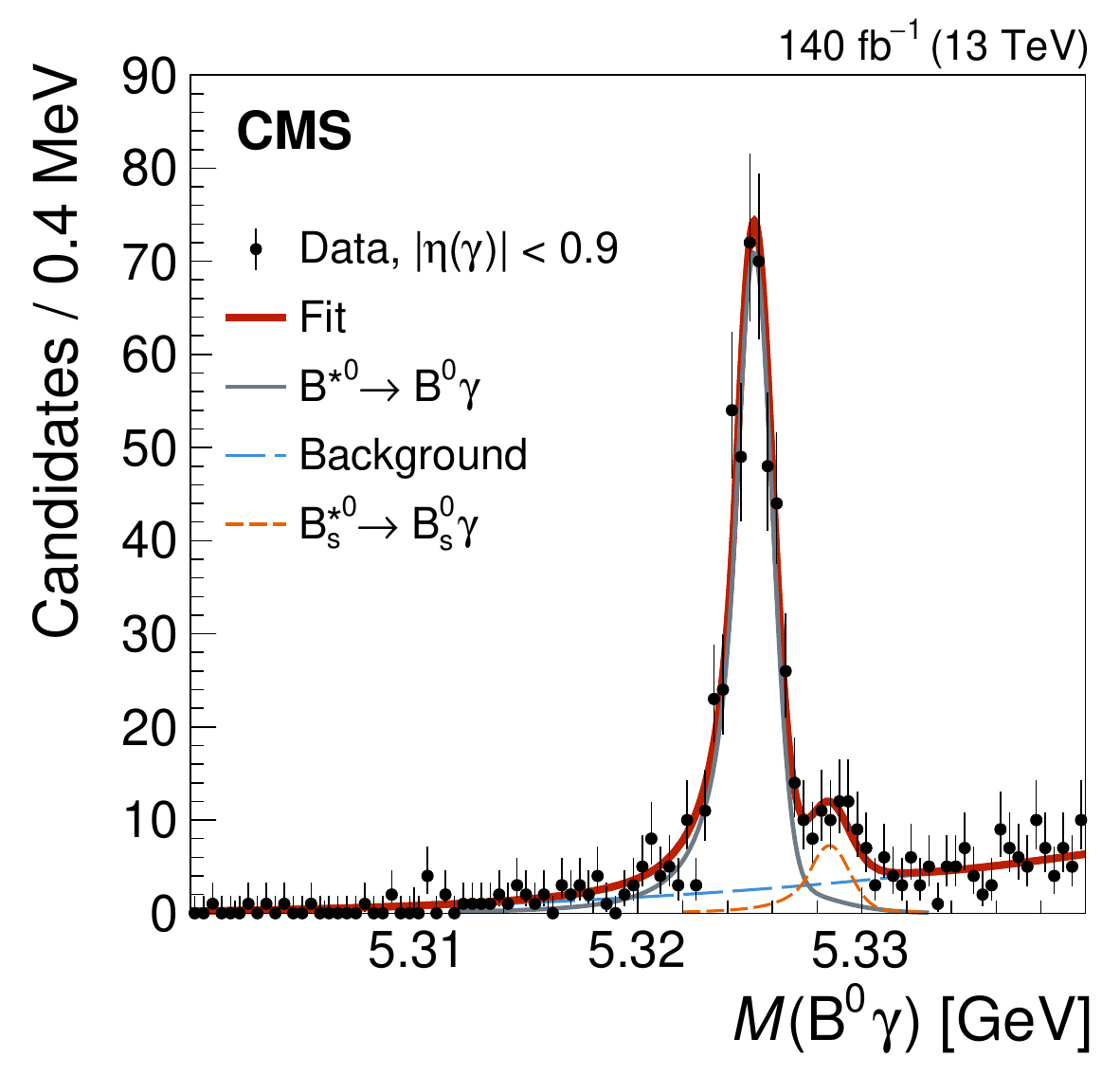}}
\caption{\textit{(Left)}  Summary of $\tau(\mathrm{B}^0)$ measurements from various experiments~\protect\cite{ATLAS3}. The result from the current analysis is shown at the top. \textit{(Right)} Distribution of the $\mathrm{B}^0\gamma$  invariant mass for photons with $|\eta|<0.9$~\protect\cite{BPH-24-011}. The total fit projection is shown with the red line.   \label{figure_b}}
\end{figure}

Additionally, a measurement of the lifetime of the B$^0$ meson, $\tau(\mathrm{B}^0)$, has been performed at ATLAS using Run~2 data (140~fb$^{-1}$)~\cite{ATLAS3}, yielding $\tau(\mathrm{B}^0) = 1.5053 \pm 0.0012\,\mathrm{(stat)}\, \pm 0.0035\,\mathrm{(syst)~}\mathrm{ps}$. It is the most precise measurement to date, as shown in Fig.~\ref{figure_b} (left). 

Finally, the first exclusive reconstruction of excited $\mathrm{B}^{*+}$, $\mathrm{B}^{*0}$, and $\mathrm{B}_\mathrm{s}^{*0}$ mesons has been conducted at CMS using Run~2 data (140~fb$^{-1}$)~\cite{BPH-24-011}. The reconstructed invariant mass of  $\mathrm{B}^{*0}\to\mathrm{B}^{0}\gamma$ is shown in Fig.~\ref{figure_b} (right). The photons are reconstructed through their conversion, $\gamma\to\mathrm{e}^+\mathrm{e}^-$, in the detector material. This analysis allows for the measurement of the hyperfine splittings, $m(\mathrm{B}^{*}) - m(\mathrm{B})$, and yields values that are one order of magnitude more precise than previous

\section{Study of all-charm tetraquarks}\label{section_tetraquarks}

\begin{figure}[h!]
\centering
\raisebox{-0.5\height}{\includegraphics[width=0.41\textwidth]{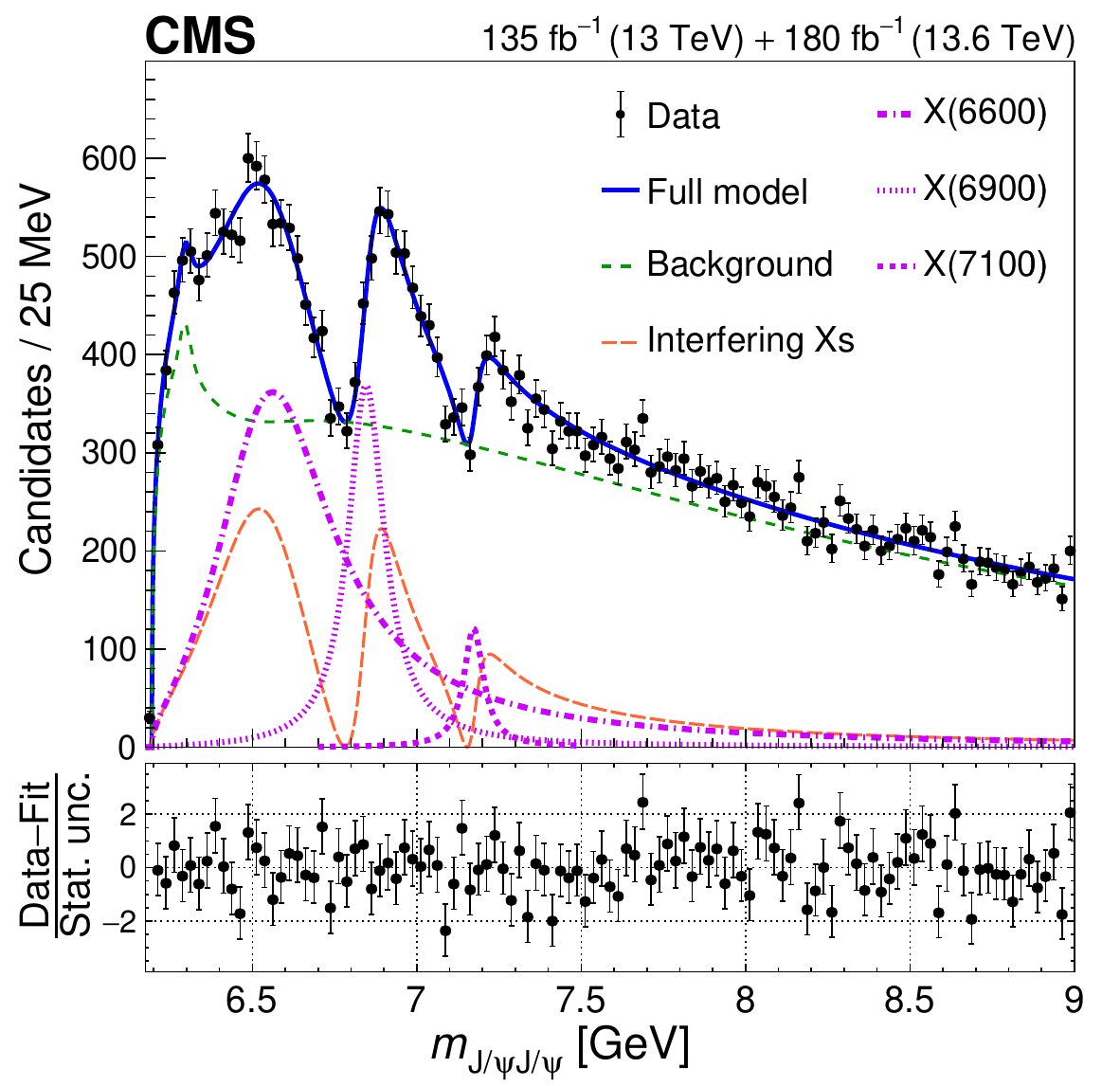}}
\hspace{1cm}
\raisebox{-0.5\height}{\includegraphics[width=0.41\textwidth]{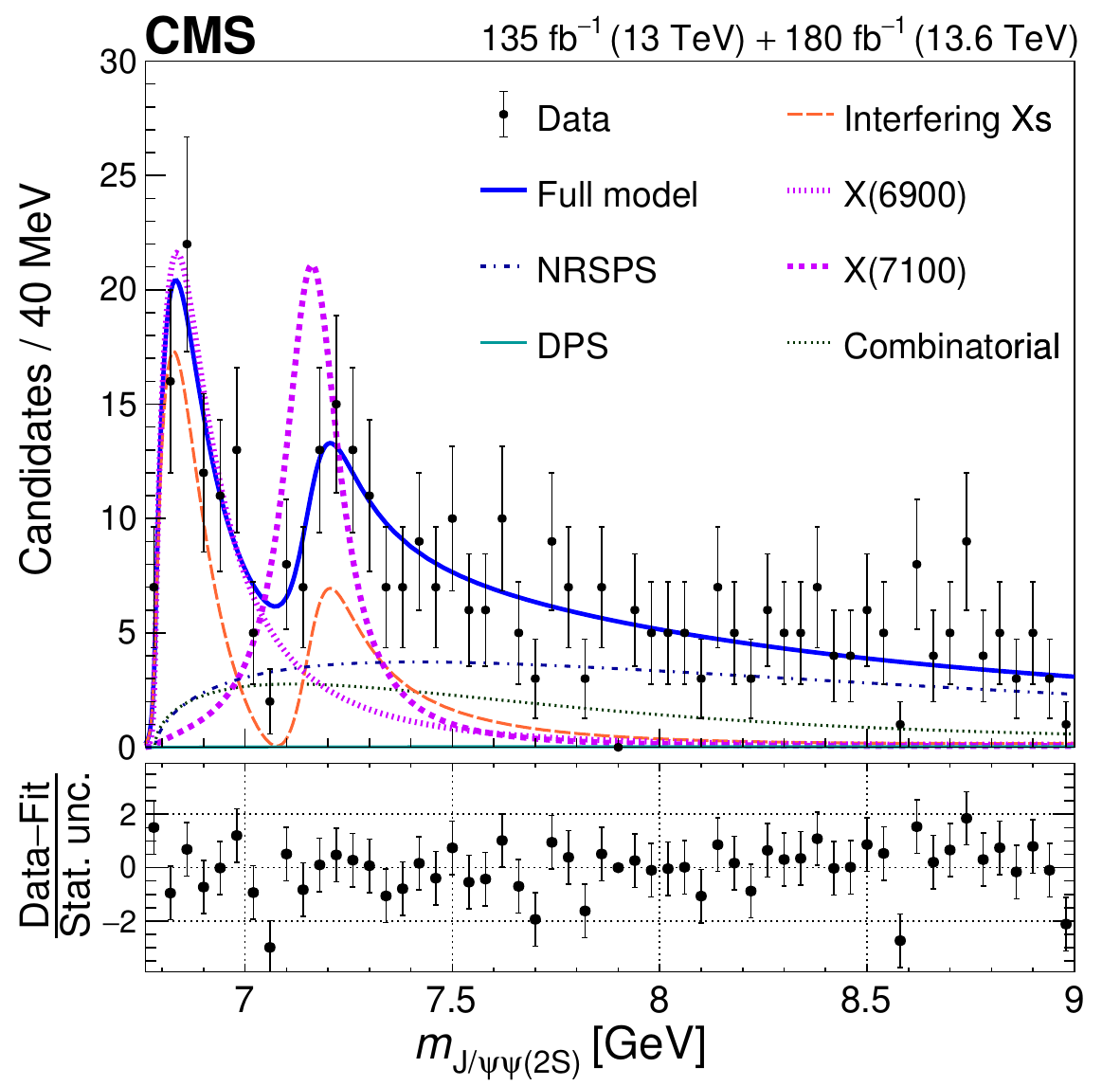}}
\caption{Invariant mass distributions in the J/$\psi$J/$\psi$~\textit{(left)} and J/$\psi\psi$(2S)~\textit{(right)} channels~\protect\cite{BPH-24-003}. The total fits are indicated with the solid blue lines.  The lower panels show the pull distributions.  \label{figure_c}}
\end{figure}

Several studies of all-charm tetraquarks have been performed at CMS with Run~2 (135~fb$^{-1}$) and Run~3 (180~fb$^{-1}$) data~\cite{BPH-24-003}. These exotic states are searched for in the double-charmonium J/$\psi$J/$\psi$ and J/$\psi\psi$(2S) mass spectra, shown in Fig.~\ref{figure_c}. The presence of three resonances, X(6600), X(6900), and X(7100), is observed in the J/$\psi$J/$\psi$ spectrum with a significance greater than 5$\sigma$. The resonance at 6.9~GeV is  also observed in the J/$\psi\psi$(2S) spectrum, while there is evidence at the level of 4$\sigma$ for a structure at 7.1~GeV in this mass spectrum.  A detailed angular analysis~\cite{CMS-Nature} in the J/$\psi$J/$\psi$ channel shows that these states are compatible with the hypothesis of  all-charm tetraquarks with $J^{\mathrm{PC}} = 2^{++}$. Similar studies were conducted at ATLAS using Run~2 data (140~fb$^{-1}$)~\cite{ATLAS1}. In particular,  the analysis in the J/$\psi\psi$(2S) channel includes for the first time the $\psi$(2S)$\to\mathrm{J}/\psi\pi^+\pi^-$ decay mode. The observation of a resonance at 6.9~GeV is confirmed, while no significant structure at 7.1~GeV is reported.

\section{Study of rare decays}\label{section_rare_decays}

Searches for the lepton flavour violating decay $\tau\to3\mu$ have been carried out at both ATLAS~\cite{ATLAS4} and CMS~\cite{BPH-24-010}. The ATLAS search is performed using Run~2 data (137~fb$^{-1}$) and is optimised for $\tau$-leptons produced in W~boson decays, while the CMS search uses Run~3 data (63~fb$^{-1}$) and is optimised for production in W, Z, and heavy-flavour decays. 
No significant excess is found in either of the searches over the background predictions. The ATLAS (CMS) analysis sets a 90\% confidence level upper limit on the branching fraction of $\mathcal{B}(\tau\to3\mu) < 8.7\times10^{-8}$ $(6.7\times10^{-8})$. 

Furthermore, CMS recently reported two new results from searches for rare decays: (i) the first observation of the $\eta\to\mu^+\mu^-\mathrm{e}^+\mathrm{e}^-$ decay~\cite{BPH-24-001} and (ii) the search for $\mathrm{B}_\mathrm{s}^0(\mathrm{B}^0)\to\mu^+\mu^-\mu^+\mu^-$ decays~\cite{BPH-24-007}. For the latter analysis,  the limits are improved by about 30\% compared with previous results, and the first simultaneous limits on the branching fractions of $\mathrm{B}_\mathrm{s}^0$ and $\mathrm{B}^0$ are provided. 

\begin{figure}[tb!]
\centering
\raisebox{-0.5\height}{\includegraphics[width=0.48\textwidth]{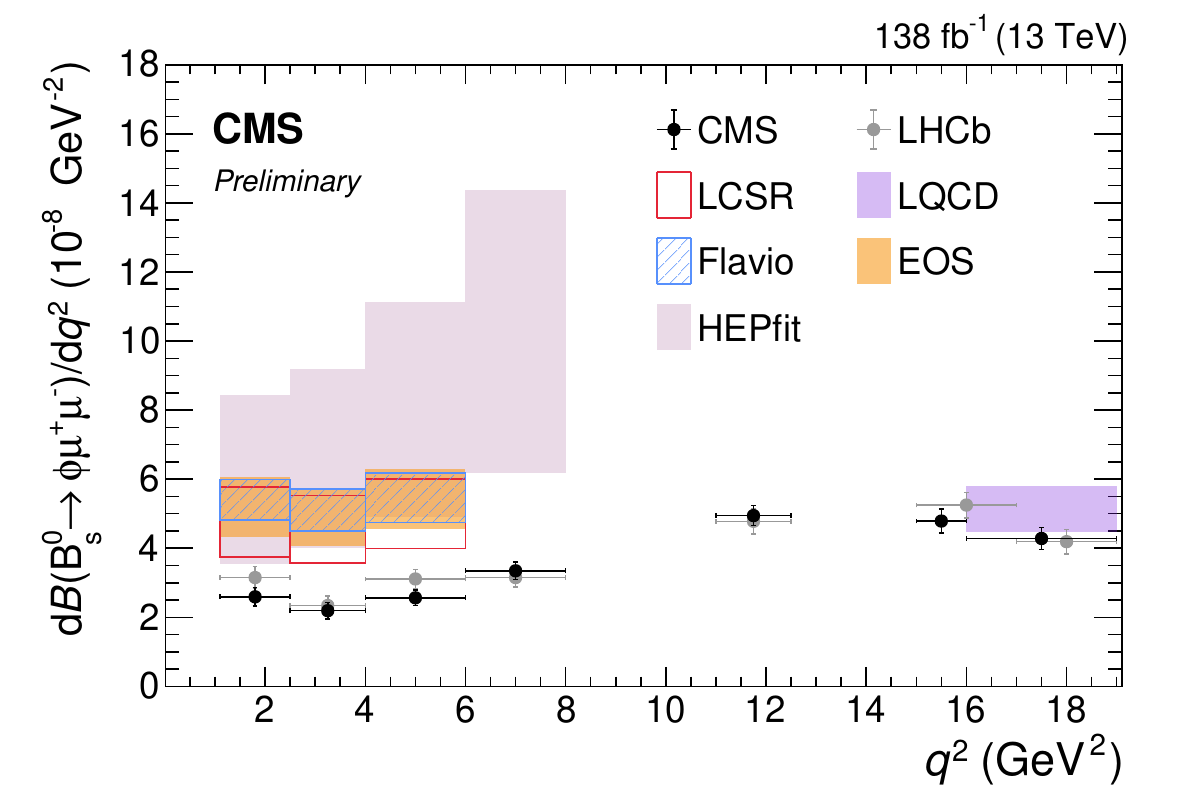}}
\hspace{0.7cm}
\raisebox{-0.5\height}{\includegraphics[width=0.44\textwidth]{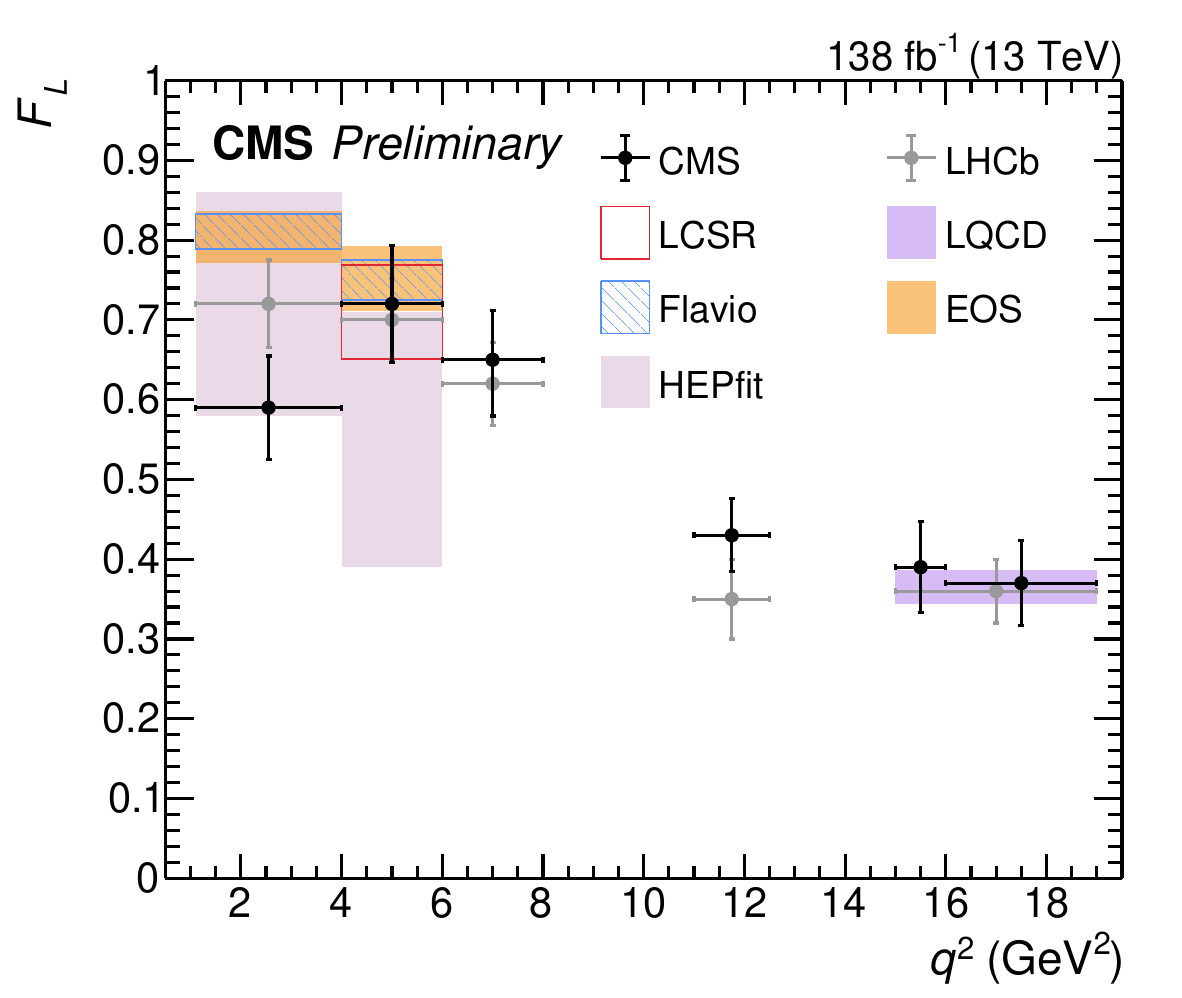}}
\caption{Differential branching fraction of the $\mathrm{B}_\mathrm{s}^0\to\phi\mu^+\mu^-$ decay~\textit{(left)} and longitudinal polarisation fraction,  $F_\mathrm{L}$, of the $\phi$ meson~\textit{(right)} as a function of $q^2$~\protect\cite{BPH-23-003}. The results are compared with several theoretical predictions. 
 \label{figure_d}}
\end{figure}

Finally, studies were performed at CMS on the rare $\mathrm{B}_\mathrm{s}^0\to\phi\mu^+\mu^-$ decay using Run~2 data (138~fb$^{-1}$)~\cite{BPH-23-003}. The analysis is motivated by tensions with the SM observed in b$\to\mathrm{s}\ell^+\ell^-$ transitions.
A differential measurement of the cross section is performed in bins of the squared dimuon invariant mass, $q^2$, in the range $1.1 < q^2 < 19$~GeV$^2$. The measurement is normalised using $\mathrm{B}_\mathrm{s}^0\to\phi\mathrm{J}/\psi(\mu^+\mu^-)$ decays. The resulting cross sections are presented in Fig.~\ref{figure_d} (left) and are in good agreement with the most recent LHCb measurement~\cite{LHCbphimumu}. The results are also compared with several theoretical predictions and show a tension at the level of up to 4.2$\sigma$. Additionally, an angular analysis has been performed.
For example, the fraction of the longitudinal polarisation of the $\phi$ mesons, $F_L$, has been measured differentially in $q^2$, as shown in Fig.~\ref{figure_d} (right). The results are found to be in good agreement with the theoretical predictions.

\section{Summary}\label{section_summary}

An overview of recent results in flavour physics using Run~2 and partial Run~3 data at the ATLAS and CMS experiments has been presented.
Thanks to their excellent trigger strategies, these detectors have sensitivity to heavy-flavour decays and provide complementary measurements to dedicated flavour physics experiments. Several key variables were measured, such as the cross sections for bottomonium and charm mesons,  the B$^0$ lifetime, and the mass differences between excited and ground B~meson states. Substantial progress has also been made in the study of all-charm tetraquarks. Finally, several analyses on rare decays significantly contribute in the search for physics beyond the SM. With the ongoing analysis of the Run~3 data samples, we look forward to a successful flavour physics programme of the ATLAS and CMS experiments. 

\section*{References}


\end{document}